\documentclass[12pt]{article}
\usepackage{latexsym}
\usepackage{epsf}

\thicklines                         
\def\footnoteitem(#1)#2{
\begin{list}{#1}{\labelwidth4.0mm \leftmargin7.0mm
\labelsep2.5mm \rightmargin7.0mm \parsep0.5ex plus0.2ex minus0.1ex
\itemsep0ex plus0.2ex }
\item #2
\end{list}
}

\def\secteq#1{ \setcounter{equation}{0}
               \renewcommand{\theequation}{#1.\arabic{equation}} }

\begin{document}
\newcommand{\be}{\begin{equation}}
\newcommand{\ee}{\end{equation}}
\newcommand{\ba}{\begin{eqnarray}}
\newcommand{\ea}{\end{eqnarray}}

\newcommand{\cL}{{\cal L}}
\newcommand{\cM}{{\cal M}}
\newcommand{\Bt}{{\tilde B}}
\newcommand{\cO}{{\cal O}}
\newcommand{\cOt}{{\tilde\cO}}
\newcommand{\bt}{{\tilde\beta}}
\newcommand{\tr}{{\mbox{tr}\,}}
\newcommand{\str}{{\mbox{str}\,}}
\newcommand{\Exp}{{\mbox{exp}\,}}
\newcommand{\Mdot}{{\dot M}}
\newcommand{\Mbar}{{M_{VS}}}
\newcommand{\tb}{{\tilde\beta}}
\newcommand{\vp}{{\vec p}}
\newcommand{\hX}{{\hat X}}
\newcommand{\diag}{{\rm diag}}
\newcommand{\sbar}{{\overline{s}}}
\newcommand{\dbar}{{\overline{d}}}
\newcommand{\ubar}{{\overline{u}}}
\newcommand{\qbar}{{\overline{q}}}
\newcommand{\psibar}{{\overline{\psi}}}
\newcommand{\ie}{{\it i.e.}}
\newcommand{\Nh}{{\hat N}}

\begin{titlepage}

\vskip 3mm

\baselineskip=20pt plus 1pt
\vskip 0.5cm

\centerline{\LARGE Effects of Quenching and Partial Quenching
on }
\vskip 0.5cm
\centerline{\LARGE Penguin Matrix Elements}
\vskip 1.0cm
\centerline{\large Maarten Golterman$^1$}
\centerline{\em  Department of Physics, Washington University,
St. Louis, MO 63130, USA$^*$}
\vskip 0.5cm
\centerline{\large Elisabetta Pallante$^2$}
\centerline{\em SISSA, Via Beirut 2--4, I-34013 Trieste, Italy}
\vskip 2.0cm
\baselineskip=12pt plus 1pt
\parindent 20pt
\centerline{\bf Abstract}
\textwidth=6.0truecm
\medskip

\frenchspacing

In the calculation of non-leptonic weak decay rates, a ``mismatch"
arises when the QCD evolution of the relevant weak hamiltonian 
down to hadronic scales is
performed in unquenched QCD, but the hadronic matrix elements
are then computed in (partially) quenched lattice QCD.  This
mismatch arises because the transformation properties of 
penguin operators under chiral symmetry change in the transition
from unquenched to (partially) quenched QCD.  Here we discuss
QCD-penguin contributions to $\Delta S=1$
matrix elements, and show that new low-energy constants 
contribute at leading order in chiral perturbation theory
in this case.  In the partially quenched case (in which
sea quarks are present), these low-energy constants are
related to electro-magnetic penguins, while in the quenched case
(with no sea quarks) no such relation exists.  As a simple
example, we give explicit results for $K^+\to\pi^+$ and
$K^0\to~vacuum$ matrix elements, and discuss the implications
for lattice determinations of $K\to\pi\pi$ amplitudes from
these matrix elements.

\nonfrenchspacing

\vskip 0.8cm
\vfill
\noindent $^*$ address after Sept. 1: \\
Dept. of Physics, San Francisco
State University, San Francisco, CA 94132, USA \\
\noindent $^1$ e-mail: {\em maarten@aapje.wustl.edu}  \\
\noindent $^2$ e-mail: {\em pallante@he.sissa.it}  \\

\end{titlepage}
\section{Introduction}
\secteq{1}
Strong and electro-magnetic penguin operators are an important part of 
the $\Delta S=1$ weak hamiltonian at hadronic scales, in particular with
respect to CP-violating kaon-decay amplitudes.  
In this paper, we will consider $LR$ operators of the form
\be
Q_{penguin}=(\sbar d)_L(\qbar X q)_R\ , \label{penguin}
\ee
where $q=u,d,s$, $X={\bf 1}$ for QCD penguins and $X=Q
=\diag(1,-1/2,-1/2)$ for electro-magnetic penguins.  
In eq.~(\ref{penguin})
\be
(\qbar_1 q_2)_{L,R}=\qbar_1\gamma_\mu P_{L,R}q_2\ , \label{currents}
\ee
 with $P_{L,R} = (1\mp \gamma_5)/2$ left- and right-handed projectors.
For each $X$ 
the color indices can be contracted in two ways, corresponding
to the operators $Q_{5,6}$ for $X={\bf 1}$ and $Q_{7,8}$
for $X=Q$~\cite{SVZ}. Of course, since the strong and
EM interactions conserve parity, $LL$ operators with $P_R\to P_L$ in the
second factor also occur; they can be written as linear combinations
of the operators $\cO_{1-4}$ (see {\it e.g.} 
ref.~\cite{dghcbtasi} for a list of all those operators).

In order to calculate the penguin contribution to non-leptonic
kaon decays, one may employ Lattice QCD in order to obtain the
non-perturbative part, while the perturbative part is encapsulated
in the Wilson coefficients, and can be calculated using 
perturbative QCD in the continuum \cite{buras}.  
The lattice part is typically done in the
quenched approximation \cite{lattice},
in which the fermion determinant is replaced by a constant.
This amounts to ignoring all sea-quark effects.
When, in the future, sea quarks will be included, one still may
wish to  use valence- and sea-quark masses which are not
equal to each other, a situation known as ``partial quenching."

Partially quenched QCD (PQQCD) can be systematically understood in a 
lagrangian framework by coupling the 
gluons to three sets of quarks \cite{bgpq}: a set of $K$
valence quarks $q_{vi}$ with masses $m_{v1},m_{v2},\dots,m_{vK}$, a set of $N$
sea quarks $q_{si}$ with masses $m_{s1},m_{s2},\dots,m_{sN}$, and a set of $K$
ghost quarks $q_{gi}$ with masses $m_{v1},m_{v2},\dots,m_{vK}$.
The ghost quarks are identical to the valence quarks, except for their
statistics, which is chosen to be bosonic \cite{morel}.  
Hence their determinant
cancels that from the valence quarks (notice they have the same 
masses), thus justifying their name: all valence-quark loops coming from
the fermion determinant are
canceled by ghost-quark loops.  Quenched QCD (QQCD) corresponds to the
special case $N=0$.  Unquenched QCD below the charm-quark threshold 
corresponds to the choice $K=N=3$ and $m_{si}=m_{vi}$, $i=1,\dots,3$. 
In this case penguin operators are written 
as in eq.~(\ref{penguin}), with $m_{v1}=m_{up}$,
$m_{v2}=m_{down}$, $m_{v3}=m_{strange}$), but the analysis of this paper 
applies for arbitrary $K$ and $N$.

The total number of quarks is thus $2K+N$, and correspondingly,
the chiral symmetry group enlarges from $SU(3)_L\times SU(3)_R$
to the graded group $SU(K+N|K)_L\times SU(K+N|K)_R$, where the
grading is a consequence of the fact that fermionic (valence or
sea) quarks can be rotated into bosonic (ghost) quarks,
and {\it vice versa} \cite{bgq,bgpq}.  A consequence of this is
that the QCD penguins (for which $X={\bf 1}$ in eq.~(\ref{penguin})) 
are no longer
right-handed singlets after making the transition from unquenched
to (partially) quenched QCD.  We will show that this leads to 
important consequences for the interpretation of
lattice results, with an argument based on chiral perturbation theory
(ChPT).  The point here is that the operators of
eq.~(\ref{penguin}) are obtained by the unquenched QCD evolution
of the weak operator from the weak scale $\sim M_W$ down to the hadronic 
scale $\sim m_c$,
so that $(\qbar q)=(\ubar u)+(\dbar d)+(\sbar s)$ is only
a singlet under flavor $SU(3)$, but not under $SU(K+N|K)$.  In contrast,
one could also imagine the situation in which strong interactions
are quenched at all scales, in which case the QCD penguins
would have taken the form
\ba
Q^{PQS}_{penguin}&=&(\sbar d)_L(\sum_{i~valence}\qbar_{vi}q_{vi}
+\sum_{i~sea}\qbar_{si}q_{si}+\sum_{i~ghost}\qbar_{gi}q_{gi})_R
\label{pqpenguin} \\
&=&\str(\Lambda\psi\psibar\gamma_\mu P_L)
\;\str(\psi\psibar\gamma_\mu P_R)\ , \nonumber \\
\Lambda_{ij}&=&\delta_{i3}\delta_{j2}\ , \label{lambda}
\ea
where $\psi=(q_v,q_s,q_g)$ and $\str$ is the supertrace, which
arises because $q_{v,s}$ and $\qbar_{v,s}$ resp.
$q_g$ and $\qbar_g$ anti-commute resp. commute.
These operators do transform as a singlet under the full PQ
symmetry group $SU(K+N|K)_R$.
They have been discussed before \cite{pwgf,gp},
but clearly, this analysis is not complete when one considers
the weak hamiltonian for which the running from the weak scale to
the hadronic scale has been calculated in unquenched QCD.

In this paper, we will consider the situation with operators of the form
(\ref{penguin}) instead of (\ref{pqpenguin}) at the hadronic scale. 
We postpone a more complete discussion,
including also $LL$ operators until later \cite{gpll},  
because the $LR$ case is somewhat simpler, and, more importantly,
because the consequences for the interpretation of (partially)
quenched lattice results are more dramatic in the $LR$ case.
\section{Penguins in (partially) quenched QCD}
\secteq{2}
The QCD penguin operators, eq.~(\ref{penguin}) with $X={\bf 1}$,
can be decomposed as 
\ba
Q^{QCD}_{penguin}&=&
\frac{K}{N}\;\str(\Lambda\psi\psibar\gamma_\mu P_L)
\;\str(\psi\psibar\gamma_\mu P_R)+
\str(\Lambda\psi\psibar\gamma_\mu P_L)
\;\str(A\psi\psibar\gamma_\mu P_R)\ , \nonumber \\
&\equiv&\frac{K}{N}\;Q^{PQS}_{penguin}+Q^{PQA}_{penguin}\ , 
\label{pqdecomp}\\
A&=&\diag(1-\frac{K}{N},\dots,1-\frac{K}{N},
-\frac{K}{N},\dots,-\frac{K}{N})\ , \label{a}
\ea
where the first $K$ (valence) 
entries of $A$ are equal to $1-{K}/{N}$,
and the next $N+K$ (sea and ghost) 
entries are equal to $-{K}/{N}$.
The superscripts $PQS$ and $PQA$ indicate that these operators transform
in the singlet and adjoint representations of $SU(K+N|K)_R$, respectively.
It is clear that these operators cannot transform into each other from
the fact that $A$ is supertrace-less, while the unit matrix is not.
It follows that in PQQCD the QCD penguin is a linear combination of
two operators which transform in different irreducible
representations (irreps) of the PQ symmetry group.

In fact, we may also embed the EM penguin $Q^{EM}_{penguin}$ 
into PQQCD, by enlarging
the charge matrix $Q$ to $Q=\diag(1,-1/2,-1/2,0,\dots,0)$.  Since $Q$
is also supertrace-less, the EM penguin is also a component of the
adjoint irrep, and $Q^{EM}_{penguin}$ and $Q^{PQA}_{penguin}$ are
thus components of the same irrep.

In the quenched case, for which $N=0$ (no sea quarks at all), the situation
is special.  The decomposition reads
\ba
Q^{QCD}_{penguin}&=&
\frac{1}{2}\;\str(\Lambda\psi\psibar\gamma_\mu P_L)
\;\str(\psi\psibar\gamma_\mu P_R)+
\str(\Lambda\psi\psibar\gamma_\mu P_L)
\;\str(\Nh\psi\psibar\gamma_\mu P_R)\ , \nonumber \\
&\equiv&\frac{1}{2}\;Q^{QS}_{penguin}+Q^{QNS}_{penguin}\ , \label{qdecomp}\\
\Nh&=&\frac{1}{2}\diag(1,\dots,1,
-1,\dots,-1)\ , \label{s}
\ea
where the first $K$ (valence) entries of $\Nh$ are equal to $\frac{1}{2}$,
and the last $K$ (ghost) entries are equal to $-\frac{1}{2}$.  The first
operator in the decomposition is a singlet, while the second is not,
under $SU(K|K)_R$
($NS$ for non-singlet).  However, the unit matrix now has a vanishing
supertrace, while $\Nh$ has not.  It is easy to show that, while 
$Q^{QS}_{penguin}$ obviously cannot transform into anything else,
$Q^{QNS}_{penguin}$ can transform into the singlet operator, so that
the non-singlet operators do not form a representation by themselves.
These group-theoretical facts correspond to the way these operators
can mix under the strong interactions: $Q^{QS}_{penguin}$ cannot mix
into any other operator, but one can easily verify that $Q^{QNS}_{penguin}$
can mix with $Q^{QS}_{penguin}$ through penguin-like diagrams.

The situation is also different with respect to the EM penguins. Since
the charge matrix
$Q$ is supertrace-less, it cannot rotate into $Q^{QNS}_{penguin}$, and
$Q^{EM}_{penguin}$ and $Q^{QNS}_{penguin}$ are not two components of
the same irrep.  Neither are $Q^{EM}_{penguin}$ and $Q^{QS}_{penguin}$,
because $Q^{QS}_{penguin}$ cannot be rotated into $Q^{EM}_{penguin}$.
We conclude that none of the three operators are related by being
members of the same irrep in the quenched case.  Note that this is
unlike the PQ case, for which both non-singlet operators are members
of the same irrep.  The difference originates in the fact that
the unit matrix is not supertrace-less for $N\ne 0$, but it is for
$N=0$.\footnote{MG thanks Noam Shoresh for instructive
discussions on this point.}
\section{Representation of penguins in ChPT}
\secteq{3}
It is well known (see {\it e.g.} ref.~\cite{dghcbtasi}) that the operators
representing QCD penguins start at order $p^2$ in ChPT, while
those representing EM penguins start at order $p^0$.\footnote{In standard
ChPT power counting EM penguins are of order $e^2p^0$; here we are
not concerned with the factor $e^2$.}  This follows
from the fact that they transform differently under $SU(3)_L\times
SU(3)_R$: QCD penguins as $(8,1)$ and EM penguins as $(8,8)$.
Denoting the adjoint representation of the PQ group by $A$, we found
in the previous section that $Q^{PQS}_{penguin}$ transform as
$(A,1)$, while $Q^{PQA}_{penguin}$ (and $Q^{EM}_{penguin}$) transform as
$(A,A)$ under $SU(K+N|K)_L\times SU(K+N|K)_R$.  
It follows that, to lowest order in ChPT and in euclidean space, 
these operators are represented by\footnote{In ref.~\cite{gp} 
we used the
shorthand notation $\alpha^8_{1,2}$ for $\alpha^{(8,1)}_{1,2}$; here we
follow the notation of ref.~\cite{betal}.}
\ba
Q^{PQS}_{penguin}&\rightarrow& -\alpha^{(8,1)}_1\;\str(\Lambda L_\mu L_\mu)
+\alpha^{(8,1)}_2\;\str(\Lambda X_+)\ , \label{chpt} \\
Q^{PQA}_{penguin}&\rightarrow&f^2\;\alpha^{(8,8)}\;
\str(\Lambda\Sigma A\Sigma^\dagger)
\ , \label{pqa} 
\ea
where
\be
L_\mu=i\Sigma\partial_\mu\Sigma^\dagger\ ,\ \ \ \ \ 
X_\pm=2B_0(\Sigma M^\dagger\pm M\Sigma^\dagger)\ ,
\label{bb}
\ee
with $M$ the quark-mass matrix, $B_0$ the parameter $B_0$ of ref.~\cite{gl},
$\Sigma=\Exp(2i\Phi/f)$ the unitary field describing the partially-quenched
Goldstone-meson multiplet, and $f$ the bare pion-decay constant
normalized such that $f_\pi=132$~MeV.  The low-energy constants (LECs)
$\alpha^{(8,1)}_{1,2}$, introduced in ref.~\cite{betal}, would also appear 
in the unquenched theory, but the appearance of the LEC
$\alpha^{(8,8)}$ is special to the PQ case.  Nevertheless, it has
a direct physical meaning, because the same LEC also appears
 in the bosonization of the EM penguin:
\be
Q^{EM}_{penguin}\rightarrow f^2\;\alpha^{(8,8)}\;
\str(\Lambda\Sigma Q\Sigma^\dagger)\ ,\label{emchpt}
\ee
because of the fact that the EM penguin $Q^{EM}_{penguin}$
and the non-singlet PQ strong penguin $Q^{PQA}_{penguin}$ are in the
same irrep of the PQ symmetry group.  In addition, taking the number
of sea quarks in the PQ theory to be the same as in the real world,
$N=3$, the LECs are the same as those of the unquenched theory
in the limit in which the $\eta'$ decouples \cite{shsh},
thus justifying their names.\footnote{This follows because the
LECs of the PQ theory only depend on $N$, but not on quark masses.}

As one would expect, the quenched case is different.  First, all LECs
are those of the $N=0$ theory, and we do not know of any
argument connecting them to those of the real world.  Second, as
pointed out in the previous section, $Q^{QNS}_{penguin}$ and
$Q^{EM}_{penguin}$ do not belong to the same irrep, and their
respective LECs are in principle different.  We therefore get the following
quenched bosonization rules, to leading order in ChPT:
\ba
Q^{QS}_{penguin}&\rightarrow& -\alpha^{(8,1)}_{q1}\;\str(\Lambda L_\mu L_\mu)
+\alpha^{(8,1)}_{q2}\;\str(\Lambda X_+)\ , \label{qchpt} \\
Q^{QNS}_{penguin}&\rightarrow& f^2\;\alpha^{NS}_q\;
\str(\Lambda\Sigma \Nh\Sigma^\dagger)
\ , \nonumber\\
Q^{EM}_{penguin}&\rightarrow &f^2\;\alpha^{(8,8)}_q\;
\str(\Lambda\Sigma Q\Sigma^\dagger)\ ,\nonumber
\ea
where the subscript $q$ indicates that these are the LECs of the quenched 
theory. One concludes that in the quenched theory yet another LEC 
$\alpha^{NS}_q$ appears, with no
counterpart in the PQ theory.  Similar new LECs will also occur for
$LL$ operators, but, as we will see in the next section, they are
particularly important in the $LR$ case, because the non-singlet $LR$ 
operators are order $p^0$ in ChPT, thus potentially leading to an 
enhancement relative to the unquenched case.
\section{$K\to\pi$ and $K\to~vacuum$ matrix elements to order $p^2$
in ChPT}
\secteq{4}
We will now show how the new LECs come into play,
by considering the simple example of $K\to\pi$ and $K\to~vacuum$
($K\to 0$) matrix elements, to leading order in ChPT.  We should
emphasize however, that the new contributions to QCD penguins found above
are properties of these operators, not only of certain matrix
elements.  In particular, new contributions would also
show up in direct $K\to\pi\pi$ matrix elements of $Q_{penguin}$
with $X={\bf 1}$.  

The new operators, $Q^{PQA}_{penguin}$ and $Q^{QNS}_{penguin}$,
do not contribute to these matrix elements at lowest order in ChPT, 
{\it i.e.} at order $p^0$.  
However, they do in general contribute at order $p^2$.  
Since  $Q^{(P)QS}_{penguin}$ starts at order $p^2$, the new
contributions from $Q^{PQA}_{penguin}$ and $Q^{QNS}_{penguin}$
compete at the {\it leading} order of the chiral expansion
of these matrix elements, 
and will have to be taken into account even if one analyzes lattice results 
using only leading-order ChPT.  

These new contributions at order $p^2$ may originate from one-loop diagrams
and from additional terms
in the bosonization of $Q^{PQA}_{penguin}$.  Those relevant for
the $K\to 0$ and $K\to\pi$ matrix elements are\footnote{One should
also consider  total-derivative terms like 
$i\partial_\mu \str(\Lambda [\Sigma A\Sigma^\dagger,L_\mu ])$
(the only one at order $p^2$),
which however does not contribute to $K\to 0$ neither to $K\to\pi$
in the mass non-degenerate case $M_K\ne M_\pi$.}
\ba
Q^{PQA}_1&=&\frac{\beta^{(8,8)}_1}{(4\pi)^2}\;
\str(\Lambda\{\Sigma A\Sigma^\dagger,L_\mu L_\mu\})\ ,
\label{nlo}\\
Q^{PQA}_2&=&\frac{\beta^{(8,8)}_2}{(4\pi)^2}\;
\str(\Lambda L_\mu\Sigma A\Sigma^\dagger L_\mu)\ ,
\nonumber \\
Q^{PQA}_3&=&\frac{\beta^{(8,8)}_3}{(4\pi)^2}\;
\str(\Lambda\{\Sigma A\Sigma^\dagger,X_+\})\ ,
\nonumber
\ea
where we introduced the $O(p^2)$ LECs $\beta^{(8,8)}_{1,2,3}$.
For the PQ $K\to\pi$ matrix element, with 
degenerate valence quark masses ($M^2=M_K^2=M_\pi^2=2B_0 m_v$),
we find at order $p^2$
\be
[K^+\to\pi^+]^{QCD}_{penguin}
=\frac{4M^2}{f^2}\Biggl\{\alpha^{(8,1)}_1-\alpha^{(8,1)}_2
-\frac{2}{(4\pi)^2}\left(1-\frac{K}{N}\right)
(\beta^{(8,8)}_1+\frac{1}{2}\beta^{(8,8)}_2+ \beta^{(8,8)}_3)
\Biggr\} \ . \label{ktopi} 
\ee
In this case the non-analytic terms coming from
eq.~(\ref{pqa}) happen to vanish, so that only contributions
from eq.~(\ref{nlo}) show up at this order, in addition to the tree-level
terms coming from eq.~(\ref{chpt}) (there are, however, chiral logarithms
coming from eq.~(\ref{pqa}) in the case $M_K\ne M_\pi$).
The $K\to 0$ matrix element with non-degenerate valence
quarks and non-degenerate sea quarks is
\ba
[K^0\to 0]^{QCD}_{penguin}
&=&\frac{4i}{f}\Biggl\{\left(\alpha^{(8,1)}_2+
\frac{2}{(4\pi)^2}\left(1-\frac{K}{N}\right)
\beta^{(8,8)}_3\right)(M_K^2-M_\pi^2)
\label{ktovac} \\
&&+\frac{\alpha^{(8,8)}}{(4\pi)^2}\left(
\sum_{i~valence}M_{3vi}^2(L(M_{3vi})-1)
-\sum_{i~valence}M_{2vi}^2(L(M_{2vi})-1)
\right.
\nonumber \\
&&\left.-\sum_{i~sea}M_{3si}^2(L(M_{3si})-1)
+\sum_{i~sea}M_{2si}^2(L(M_{2si})-1)\right)\Biggr\} \ .\nonumber
\ea
Here
\be
L(M)=\log{\frac{M^2}{\Lambda^2}}\ ,\label{log}
\ee
and the result is given in the $\overline{MS}$ scheme, with $\Lambda$
the running scale.  $M_{3si}$ ($M_{2si}$) is the mass of a meson made out 
of the 3rd (2nd) valence (\ie\ the strange (down)) quark and the $i$th 
sea quark; $M_{3vi}$ ($M_{2vi}$) is the mass of a meson made out of
the 3rd (2nd) valence (\ie\ the strange (down)) quark and the $i$th valence 
quark.  At the order we are working, $M_{3si}^2-M_{2si}^2=
M_{3vi}^2-M_{2vi}^2=M_K^2-M_\pi^2$, which follows from
$M_{3si}^2=B_0(m_{v3}+m_{si})$, {\it etc.}

{}From these results we learn several things.  First, the contributions
from the new operators to eqs.~(\ref{ktopi}) and (\ref{ktovac})
indeed appear at order $p^2$, \ie\ the same order as the
leading order in the unquenched theory. Notice that in general, as 
{\it e.g.}
in eq.~(\ref{ktovac}), they contribute to a weak matrix element at leading 
order with non-analytic terms of the form
$M^2\log{M^2}$, which are absent in the 
unquenched case. Second, as one would expect,
these results also contain the unquenched result as a particular case.  
This can be seen
by choosing the number of sea quarks equal to the number of valence
quarks (\ie\ $N=K=3$), and by equating corresponding quark masses,
$m_{si}=m_{vi}$, $i=1,\dots,K$.  For this choice, the terms proportional
to $\alpha^{(8,8)}$ and $\beta^{(8,8)}_{1,2,3}$
in both matrix elements vanish, as they should.
That this is also true 
at higher orders, as well as for other matrix elements, can be deduced from a
quark-flow argument.  For $N=K$, the first $K$ entries in $A$ 
(eq.~(\ref{a})) vanish, leaving only sea and ghost quarks in 
the second factor ($\str(A\psi\psibar\gamma_\mu P_R)$) of
$Q^{PQA}_{penguin}$.  If there are only valence
quarks on the external lines, these sea and ghost quarks have to 
produce loops, which cancel if $N=K$ and their masses are pairwise
equal.  Note that the choice $N=K$ is necessary, because only in that case
does the number of quarks in $(\qbar Xq)_R$ in eq.~(\ref{penguin})
correspond to the number of sea quarks.

At this point it is interesting to note that, although the terms
proportional to $\alpha^{(8,8)}$ and $\beta^{(8,8)}_{1,2,3}$
are an unexpected ``contamination,"
they still contain physical information about EM penguin matrix 
elements.  There is also a way to avoid this contamination, by
considering instead $Q^{PQS}_{penguin}$ alone. In practice,
this implies throwing out all Wick contractions in which $q$ and
$\qbar$ in eq.~(\ref{penguin}) (with $X={\bf 1}$) are contracted, except 
when they correspond to sea quarks.

Quenched ($N=0$) results are obtained by replacing
$\alpha^{(8,1)}_{1,2}\to\alpha^{(8,1)}_{q1,2}$,
$\alpha^{(8,8)}\to\alpha^{NS}_q$ and
$\beta^{(8,8)}_{1,2,3}\to\beta^{NS}_{q1,2,3}$ in 
eqs.~(\ref{ktopi},\ref{ktovac}),
and by dropping all terms containing sea quarks.  One obtains
\ba
[K^+\to\pi^+]^{QCD}_{penguin}
&=&\frac{4M^2}{f^2}\Biggl\{\alpha^{(8,1)}_{q1}-\alpha^{(8,1)}_{q2}
-\frac{1}{(4\pi)^2}
(\beta^{NS}_{q1} +\frac{1}{2}\beta^{NS}_{q2}+\beta^{NS}_{q3})
\Biggr\} \ , \label{qktopi}\\
{[}K^0\to 0{]}^{QCD}_{penguin}
&=&\frac{4i}{f}\Biggl\{\left(\alpha^{(8,1)}_{q2}+
\frac{1}{(4\pi)^2}
\beta^{NS}_{q3}\right)(M_K^2-M_\pi^2)
\label{qktovac} \\
&&\hspace{-1.2truecm}+\frac{\alpha^{NS}_q}{(4\pi)^2}\left(
\sum_{i~valence}M_{3vi}^2(L(M_{3vi})-1)
-\sum_{i~valence}M_{2vi}^2(L(M_{2vi})-1)
\right)
\Biggr\} \ .\nonumber
\ea
In this case the terms proportional to $\alpha^{NS}_q$ 
and $\beta^{NS}_{q1,2,3}$ are a genuine contamination 
of the tree-level results coming from $Q^{QS}_{penguin}$, and do not carry
any physical information about EM penguins.
Hence there is no reason to expect that a quenched lattice computation
of a $Q_5$ or $Q_6$ matrix element has anything to do with the
real world.  Again, one may consider only $Q^{QS}_{penguin}$ in order
to determine $\alpha_{q1,2}^{(8,1)}$.
This would mean that no ``eye graphs" with $q$ and $\qbar$ contracted 
would be considered
at all.\footnote{The remaining ``eye graphs" are those for which 
$q$ ($\qbar$) is 
contracted with $\bar{s}$ ($d$) in eq.~(\ref{penguin}).}

We conclude this section with the EM contributions, up to order $p^2$,
to the matrix elements considered here (for the unquenched case see also 
ref.~\cite{CG}).  For the $K\to\pi$ matrix
element with degenerate valence quarks and degenerate sea quarks
we obtain from eq.~(\ref{emchpt}) 
\be
[K^+\to\pi^+]^{EM}_{penguin}=6\alpha^{(8,8)}\Biggl\{1
-\frac{2N}{(4\pi f)^2}M_{VS}^2(L(M_{VS})-1)
\Biggr\}\ ,\label{emktopi}
\ee
where $M_{VS}^2=B_0(m_v+m_s)$ is the (tree-level) mass-squared
of a meson made out of one valence and one sea quark,
and for the $K\to 0$ matrix element
\ba
[K^0\to 0]^{EM}_{penguin}&=&-2if\frac{\alpha^{(8,8)}}
{(4\pi f)^2}\Bigg\{M_\pi^2 L(M_\pi)-2M_K^2 L(M_K)+M_{33}^2 L(M_{33})
\nonumber \\
&&+\sum_{i~sea}\left(M_{2si}^2(L(M_{2si})-1)-M_{3si}^2(L(M_{3si})-1)\right)
\Biggr\}\ .\label{emktovac}
\ea
The quenched result is obtained by dropping all terms containing
the sea-quark mass, and by replacing $\alpha^{(8,8)}
\to\alpha^{(8,8)}_q$.

In addition to the operators of eq.~(\ref{nlo}) with $A$ replaced
by $Q$, there are two more operators which contribute 
counterterms to eqs.~(\ref{emktopi},\ref{emktovac}) at this order:%
\footnote{ The total-derivative term 
$i\partial_\mu \str(\Lambda [\Sigma Q\Sigma^\dagger,L_\mu ])$
gives a contribution to $K\to\pi$
in the mass non-degenerate case $M_K\ne M_\pi$.}
\ba
Q^{PQA}_4&=&\frac{\beta^{(8,8)}_4}{(4\pi)^2}\;
\str(\Lambda[\Sigma Q\Sigma^\dagger,X_-])\ , \label{morenlo}\\
Q^{PQA}_5&=&\frac{\beta^{(8,8)}_5}{(4\pi)^2}\;
\str(\Lambda\Sigma Q\Sigma^\dagger)\;\str(X_+)\ .\nonumber
\ea
The EM counterterm contributions are
\ba
[K^+\to\pi^+]^{EM}_{c.t.}&=&
\frac{4}{(4\pi f)^2}\Biggl((\beta^{(8,8)}_1 - \beta^{(8,8)}_2 
+7\beta^{(8,8)}_3
-6\beta^{(8,8)}_4)M^2+3\beta^{(8,8)}_5NM_{SS}^2
\nonumber \\
&&-24\alpha^{(8,8)}(\lambda_5M^2+
\lambda_4NM_{SS}^2)
\Biggr) ,
\label{emct} \\
{[}K^0\to 0{]}^{EM}_{c.t.}&=&
-\frac{4if}{(4\pi f)^2}\beta^{(8,8)}_3(M_K^2-M_\pi^2)\ ,\nonumber
\ea
where the strong $O(p^4)$ LECs $\lambda_i$ enter via wave-function 
renormalization and are related to the
Gasser--Leutwyler $L_i$ \cite{gl} by
\be
\lambda_i=16\pi^2L_i\ .\label{lambdas}
\ee
Quenched results follow by setting $N=0$ and replacing
$\beta^{(8,8)}_i\to\beta^{(8,8)}_{qi}$, $i=1,2,3,4,5$,
$\lambda_k\to\lambda_{qk}$, $k=4,5$.
\section{Conclusion}
\secteq{5}
In this paper we have considered the question as to what happens when one
evaluates the matrix elements of QCD penguin operators in the quenched
or partially quenched approximations which are commonly used in
Lattice QCD.  QCD penguins for $\Delta S=1$ weak operators 
transform as an octet under $SU(3)_L$ and as a singlet under
$SU(3)_R$.  However, once one makes the transition from unquenched
QCD, in which the weak hamiltonian at hadronic scales $\sim m_c$ 
is calculated,
to quenched or partially quenched QCD, in which the lattice computations
are done, the chiral group $SU(3)_L\times SU(3)_R$ enlarges 
to $SU(K+N|K)_L\times SU(K+N|K)_R$,
and the corresponding statement is no longer true.  

In the simple case of $LR$ penguins considered here 
({\it cf.} eq.~(\ref{penguin})), the QCD
penguins are a linear combination of two operators which transform
in different irreps of the (partially) quenched symmetry group.
The first of these transforms as a singlet under $SU(K+N|K)_R$,
much like in the unquenched case, and has a similar chiral behavior,
with matrix elements linear in the quark masses at leading
order in ChPT.  The second operator, however, transforms in the
adjoint representation (which generalizes the octet representation
of $SU(3)$) of {\it both} $SU(K+N|K)_L$ and $SU(K+N|K)_R$.
This new operator does not contribute at the lowest order
in the chiral expansion to the matrix elements of interest, but it does 
contribute at next-to-leading order.  
However, because of the fact that this operator starts at
order $p^0$ in (partially) quenched
ChPT, its leading non-vanishing contributions compete with the 
leading contributions of the operators already present in the
unquenched case.

At leading order in ChPT, the new operator corresponds to one new
low-energy constant, $\alpha^{(8,8)}$ in the partially quenched case 
({\it cf.} eq.~(\ref{pqa})) and $\alpha_q^{NS}$ in the quenched case 
({\it cf.} eq.~(\ref{qchpt})), 
while at higher orders new LECs proliferate, as usual.
In the PQ case, these LECs turn out to be those corresponding to the
EM penguin, because the new non-singlet operator and the EM penguin
transform in the same irreducible representation 
of the enlarged chiral-symmetry group.  In
the quenched case, with no sea-quarks at all, this is not the case,
and the new operators must be considered a pure quenching artifact.
We have demonstrated the way the new operators work with the simple
examples of $K\to~vacuum$ and $K\to\pi$ matrix elements; leading-order
expressions useful for practical applications can be found in
section 4. 

The implications of our analysis for the use of $K\to\pi$ and $K\to 0$
matrix elements in the determination of $K\to\pi\pi$ amplitudes are
the following.  To leading order in ChPT, the physical $K\to\pi\pi$ 
amplitude is determined by $\alpha_1^{(8,1)}$, and therefore this is the
LEC one wishes to extract from the $K\to\pi$ matrix element.
It is then clear that the way to do this is to consider only the
singlet penguin, $Q_{penguin}^{PQS}$, in the PQ theory with $N=3$
light sea quarks.  As already mentioned in section 4, this is
equivalent to omitting all Wick contractions in which $q$ and
$\qbar$ in eq.~(\ref{penguin}) (with $X={\bf 1}$)
are contracted, except when they
correspond to sea quarks.  Of course, since the $K\to\pi$ 
matrix element only gives the linear combination $\alpha_1^{(8,1)}
-\alpha_2^{(8,1)}$, the $K\to 0$ matrix element of 
$Q_{penguin}^{PQS}$ is also needed, as usual \cite{betal}.

In the case that $N\ne 3$, one can still follow a similar strategy
in order to determine $\alpha_1^{(8,1)}$.  However, since the LECs
depend on the number of light sea quarks $N$, there is no reason
why the result should be the same as that of the real world,
which has $N=3$. In the quenched case ($N=0$), the same strategy 
corresponds to considering only $Q_{penguin}^{QS}$, or equivalently, no
diagrams with $q$ and $\qbar$ contracted at all. This would be 
the most reasonable strategy to adopt if one assumes that 
$\alpha_1^{(8,1)}(N=0)\approx\alpha_1^{(8,1)}(N=3)$. Another possibility, 
in the partially quenched case with $N\neq 3$ and in the quenched case,
is to include both the singlet and non-singlet operators in the determination 
of $K\to\pi$ and $K\to vacuum$ matrix elements, and also in the 
conversion to $K\to\pi\pi$ amplitudes \cite{gpll}. 
If one does
include $q\qbar$ contractions for the valence quarks, the results
will depend on the new LECs corresponding to the operators
$Q_{penguin}^{PQA}$ of the partially quenched case or $Q_{penguin}^{QNS}$ 
of the quenched case and $Q_{1,2,3}$ of eq.~(\ref{nlo}).  
With either strategy, one cannot expect, {\em a priori},
to obtain a realistic result for the penguin
contributions to $K\to\pi\pi$ amplitudes. In addition, since it is not known 
which of these strategies is a better approximation, the difference between 
them can be interpreted as an indication of the quenching error.

For $LL$ penguin operators a similar analysis applies.  When one
replaces the second factor in eq.~(\ref{penguin}) by a left-handed
current, $(\qbar Xq)_R\to (\qbar Xq)_L$, this factor is again not
a singlet under $SU(K+N|K)_L$, and new operators will appear.
In this case however, we expect the new operator to start at order $p^2$
in ChPT.  Since the lowest-order operators 
will again not contribute at tree level
to matrix elements of interest, the new $LL$ operators will only be 
relevant for a next-to-leading order analysis of lattice matrix elements
\cite{gpll}.

Finally, we should emphasize that the effects of
(partial) quenching on penguin operators discussed here are a
property of the penguin operators themselves and not only of 
certain matrix elements, and will therefore have
similar consequences for any weak matrix element to which such
operators contribute, including not only non-leptonic kaon decays,
but also non-leptonic $B$ decays.

\subsubsection*{Acknowledgements}
We would like to thank Tom Blum, Norman Christ, Bob Mawhinney, Amarjit 
Soni and other members of the RBC collaboration for asking questions leading
to this paper.  We also thank Claude Bernard, Santi Peris and Steve Sharpe
for helpful discussions.
MG would like to thank the particle theory groups of SISSA (Trieste, Italy)
and the Universitat Aut\`onoma (Barcelona, Spain) for hospitality.
MG is supported in part by the US Dept. of Energy.

\end{document}